\documentclass[%
 reprint,
superscriptaddress,
 amsmath,amssymb,
 aps,
]{revtex4-2}

\usepackage{graphicx}% Include figure files
\usepackage{dcolumn}% Align table columns on decimal point
\usepackage{bm}% bold math
\usepackage{comment}

\begin{document}

\preprint{APS/123-QED}

\title{Non-magnetic Fractional Conductance in High Mobility InAs Quantum Point Contacts}

\author{I. Villar Rodriguez}
  \email{uceeivi@ucl.ac.uk}
\author{Y. Gul}%
\affiliation{%
 London Centre for Nanotechnology, University College London, 17-19 Gordon Street, London WC1H 0AH, United Kingdom\\
}%
\affiliation{%
 Department of Electronic and Electrical Engineering, University College London, Torrington Place, London WC1E 7JE, United Kingdom
}%

\author{C.P. Dempsey}
\affiliation{
 Department of Electrical and Computer Engineering, University of California, Santa Barbara, CA 93106, USA
}%
\author{J.T. Dong}
\affiliation{%
 Materials Department, University of California, Santa Barbara, CA 93106, USA
}%

\author{S.N. Holmes}
\affiliation{%
 Department of Electronic and Electrical Engineering, University College London, Torrington Place, London WC1E 7JE, United Kingdom
}%

\author{C.J. Palmstrøm}
\affiliation{
 Department of Electrical and Computer Engineering, University of California, Santa Barbara, CA 93106, USA
}%

\affiliation{%
 Materials Department, University of California, Santa Barbara, CA 93106, USA
}%

\author{M. Pepper}
\affiliation{%
 London Centre for Nanotechnology, University College London, 17-19 Gordon Street, London WC1H 0AH, United Kingdom\\
}%
\affiliation{%
 Department of Electronic and Electrical Engineering, University College London, Torrington Place, London WC1E 7JE, United Kingdom
}%

\date{\today}% It is always \today, today,

\begin{abstract}
In this letter, we report the magneto-electronic properties of high mobility InAs quantum point contacts grown on InP substrates. The 1D conductance reaches a maximum value of 17 plateaus, quantized in units of 2e$^2$/h, where $e$ is the fundamental unit of charge and $h$ is Planck’s constant. The in-plane effective g-factor was estimated to be $- 10.9 \pm 1.5$ for subband $N=1$ and $- 10.8 \pm 1.6$ for subband $N=2$. Furthermore, a study of the non-magnetic fractional conductance states at 0.2(e$^2$/h) and 0.1(e$^2$/h) is provided. While their origin remains under discussion, evidence suggests that they arise from strong electron-electron interactions and momentum-conserving backscattering between electrons in two distinct channels within the 1D region. This phenomenon may also be interpreted as an entanglement between the two channel directions facilitated by momentum-conserving backscattering.
\end{abstract}

\maketitle

%\tableofcontents

\section{\label{sec:level1}Introduction}
Quantization of conductance in units of 2e$^2$/h in one-dimensional (1D) quantum point contacts (QPCs) was first observed by D.A. Wharam \cite{Wharam} and B.J. van Wees \cite{Wees} when investigating a gate-controlled channel defined in the two-dimensional electron gas (2DEG) of a GaAs/AlGaAs heterostructure. Since then, many reports have studied the interesting physics of these constrictions: the effect of a source-drain bias voltage \cite{Patel,TMChen1,TMChen2}, the appearance of the 0.7 conductance anomaly \cite{Thomas0.71,Thomas0.72,Thomas0.73,Kristensen0.7} and other many-body effects \cite{Hew,Smith,KumarMany}. One of the most highlighted recent advances in the field demonstrated the potential to spatially manipulate electron spins in a QPC \cite{TEF1,TEF2}. This is a crucial component for spin-based technologies. Additionally, the realization of an all-electrical spin field-effect transistor was achieved through the implementation of two QPCs \cite{FET}.

1D quantization of conductance is observed in ballistic constrictions where the mean free path of the electrons is larger than the size of the split gates. Therefore, high quality heterostructures with high 2DEG electron mobility are required. Furthermore, a 2DEG with strong spin-orbit coupling (SOC) would facilitate the manipulation of the electron spins. A material that meets all these requirements would be an ideal candidate for spintronic and quantum information devices.

High electron mobility InAs quantum wells (QWs) have been grown in both lattice-matched GaSb \cite{Shojaei-GaSb,Thomas-GaSb} and lattice-mismatched InP \cite{InAs1Dwire3, ManfraInP} substrates. However, the reported presence of side wall conduction in InAs QWs grown on GaSb substrates \cite{Sidewall} makes InP a better choice. Moreover, recent advancements \cite{connor} in wafer manufacturing have improved upon the current state-of-the-art QW thickness $t_{QW}$ ($t_{QW}=4$ nm for In$_{0.75}$Al$_{0.25}$As \cite{ManfraInP} barrier layers and $t_{QW}=7$ nm for
In$_{0.81}$Al$_{0.19}$As \cite{7nmQW} barrier layers) of InAs QWs grown on InP substrates. This advancement shows an improvement in the highest achievable mobility.  Other significant properties of InAs include its small effective mass \cite{effmass} and its large bulk g-factor. For comparison, the g-factor of bulk InAs is - 15 \cite{InAsgfactor}, which is considerably larger than that of InGaAs (- 9) \cite{Holmes_InGaAs-gfactor} or GaAs (- 0.44) \cite{GaAsgfactor}. Furthermore, InAs exhibits strong SOC that allows efficient control of spin currents. The large g-factor of InAs leads to Zeeman splitting at low magnetic fields, which in proximity to an s-wave superconductor could introduce Majorana zero modes \cite{InAsmajo3,InAsmajo4,InAsmajo5,microsoft-majo} that can be utilized for topological quantum computing applications \cite{majo-topo}.
Several studies have already been conducted on 1D InAs QPCs \cite{InAs1Dwire1,InAs1Dwire2, InAs1Dwire3, InAs1Dwire4}, with a maximum of eight pronounced quantized conductance plateaus observed at 1.5 K \cite{InAscleanwire}.

Strong electron-electron interactions occur when the carrier density in the QPC is low. In this state, non-magnetic fractional quantization of conductance has previously been reported in strained-Ge \cite{Yilmazpaper}, InGaAs \cite{Leipaper} and GaAs \cite{Sanjeevpaper} QWs. Although the origin of these new quantum states is not fully understood, their study could lead to future applications in quantum computing.

In this work, we present the magneto-transport properties of gate-defined 1D QPCs fabricated on InAs QWs grown on InP substrates, with QW thicknesses exceeding those reported in earlier studies. We demonstrate clean and controllable  1D behavior. Furthermore, strong electron-electron interactions at low carrier densities are investigated. Here, non-magnetic fractional quantization of conductance is observed, and possible effects leading to this phenomenon are discussed.

\section{Device fabrication and experimental setup} \label{fab&setup}
The device presented in this work is an InAs QW heterostructure grown on a semi-insulating Fe-doped InP (001) substrate by molecular beam epitaxy (MBE). The 2DEG lies 130 nm below the surface and is surrounded by In$_{0.72}$Ga$_{0.28}$As cladding layers. The width of the QW is 16 nm, exceeding the maximum previously achievable width for InAs QWs grown on InP substrates \cite{ManfraInP,7nmQW}. A more detailed explanation of the growth conditions can be found in Ref. \cite{connor}.

For 1D electrical and magneto-transport measurements, Hall bars of 1400 $\mu$m in length and 80 $\mu$m in width were fabricated. Using a mixture of H$_2$SO$_4$:H$_2$O$_2$:H$_2$O (1:8:120), the Hall bar mesa was etched so that the QPC channel lies along the [1$\overline{1}$0] direction. AuGeNi Ohmic contacts connecting the 2DEG and the surface were evaporated onto the sample and annealed at 450 $^o$C for 2 minutes. To insulate the 2DEG from the gates, a 35 nm Al$_2$O$_3$ dielectric layer was deposited using atomic layer deposition (ALD). Windows for the Ohmic contacts were opened by etching through the dielectric. Finally, the split gates and top gates were patterned onto the wafer using electron beam lithography (EBL). Cross-linked polymethyl methacrylate (PMMA) was used as a dielectric between the split gates and the top gates, and Ti/Au was deposited to form the gates. The results presented in this letter were obtained from split gates measuring 500 nm in length and width or 400 nm in length and 500 nm in width. At 1.5 K, the 2DEG carrier density is 3.6 $\times$ 10$^{11}$ cm$^{-2}$, and the electron mobility is 9.3 $\times$ 10$^5$ cm$^2$/Vs. The effective electron mass and the Rashba coefficient were determined as (0.031 $\pm$ 0.002)$m_e$, where $m_e$ is the rest electron mass, and 12 $\times$ 10$^{-12}$ eVm , respectively \cite{connor}.

\begin{figure}[t]
\includegraphics[width=.5\textwidth]{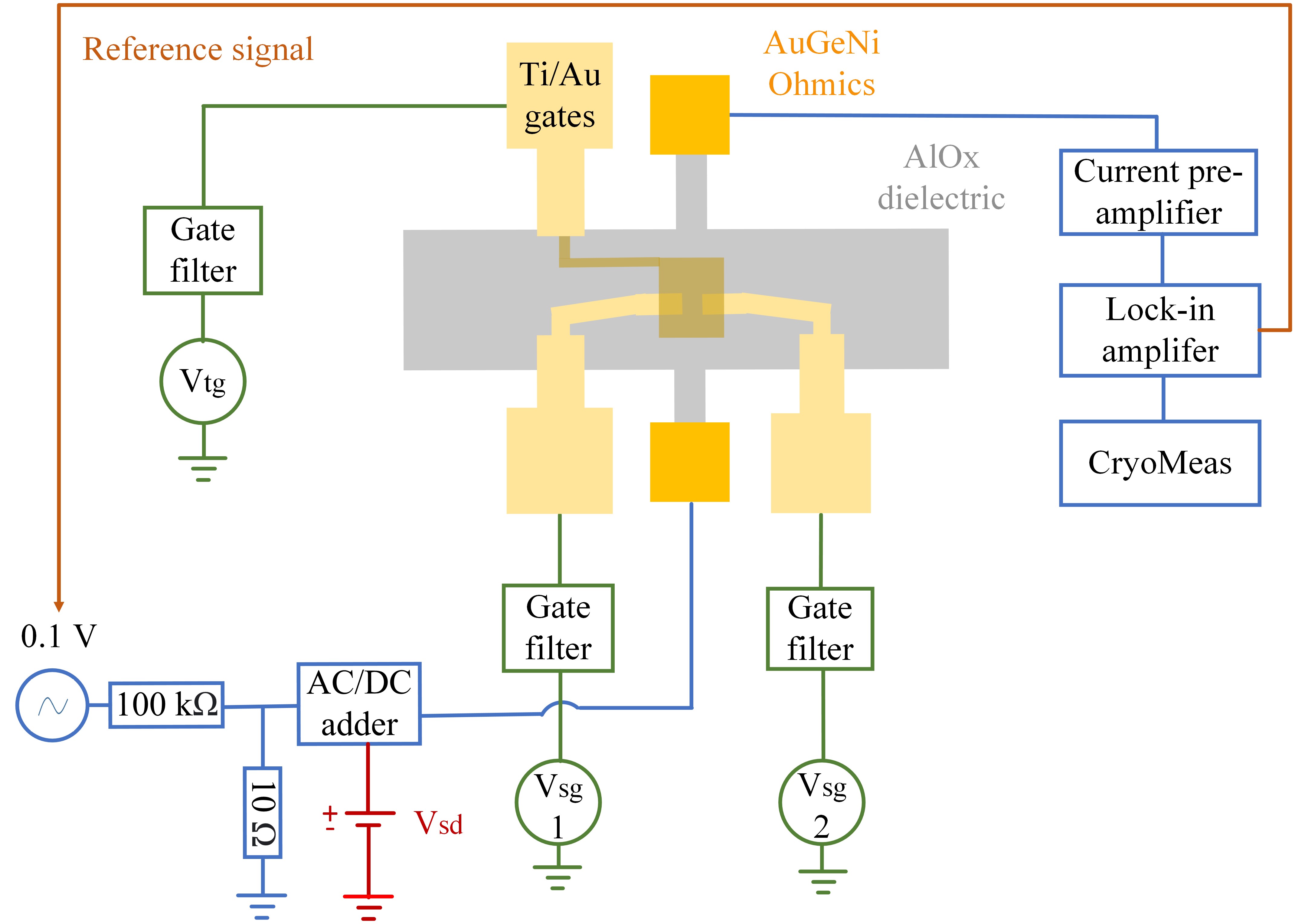}
\caption{\label{setup} Two-terminal conductance measurement setup. A constant AC differential voltage drop across the device is applied using a potential divider. A DC source-drain bias voltage, $V_{sd}$, can be superimposed onto the signal using a AC/DC adder. All gates are protected by a low pass gate filter. The outgoing signal is amplified by a current pre-amplifier and measured by a lock-in amplifier. CryoMeas \cite{cryomeas} is the program utilized to record the data.}
\end{figure}

In all experiments, the two-terminal differential conductance ($G$) was measured as a function of split gate voltage ($V_{sg}$) at a constant top gate voltage ($V_{tg}$), while a 10 $\mu$V AC voltage at 33 Hz was applied. The measurements were taken in a cryo-free dilution refrigerator at 220 mK over multiple cooldowns. The measurement configuration is shown in Fig. \ref{setup}.

\section{Results and Discussion} \label{results}
\subsection{1D conductance quantization}
\begin{figure}[h!]
    %\centering
    \includegraphics[width=.5\textwidth]{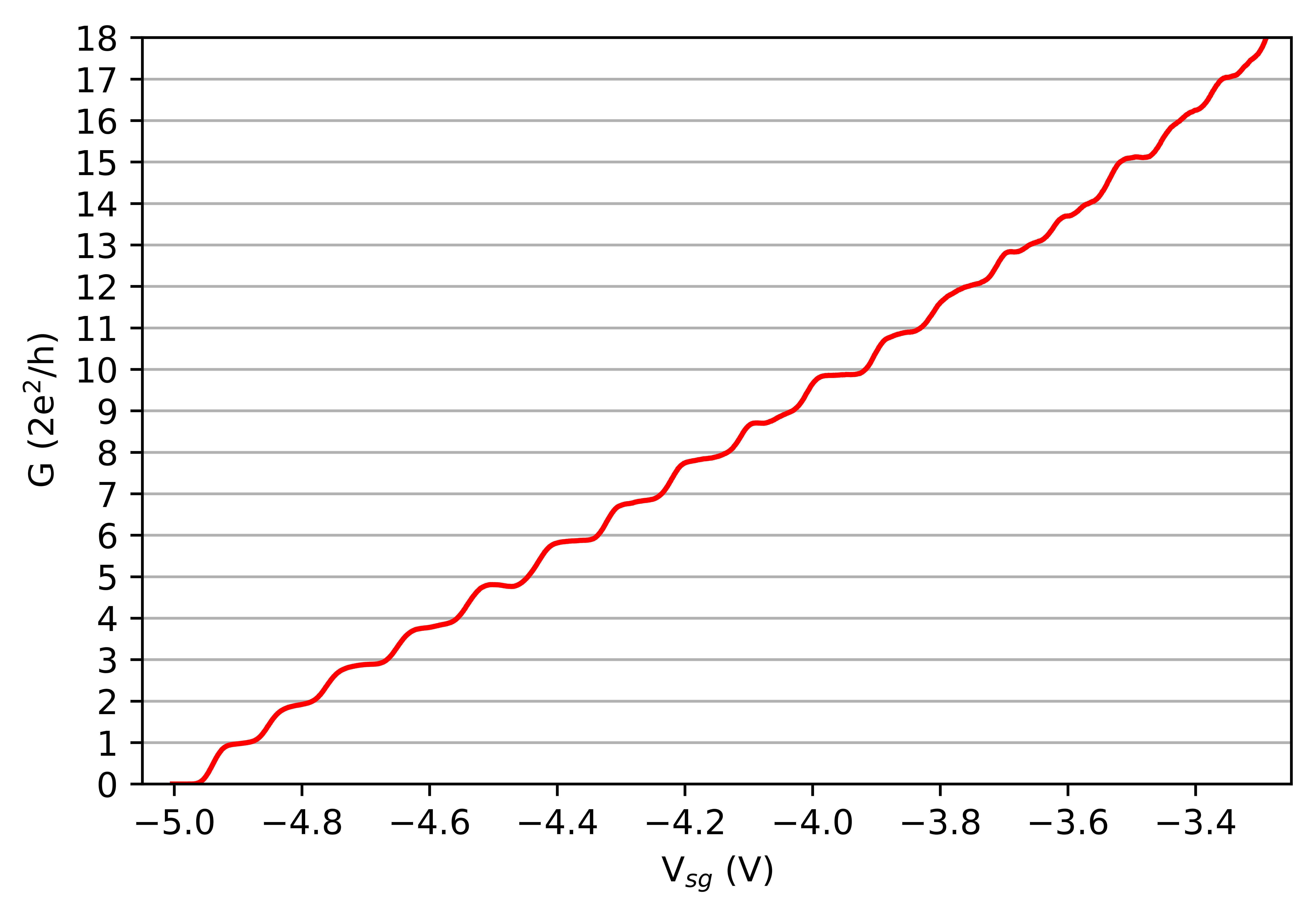} %[width=1\textwidth]
    \caption{$G$ as a function of $V_{sg}$ when $V_{tg}=0$ V. The split gate dimensions are 400 nm in length and 500 nm in width, and up to seventeen plateaus quantized in units of 2e$^2$/h can be observed. The series resistance for this measurement was 1.3 k$\Omega$, determined individually for each cooldown and each QPC.}
    \label{singletrace}
\end{figure}

Fig. \ref{singletrace} illustrates the quantized 1D conductance as the split gate voltage is raised from the pinch-off state. We observe seventeen plateaus quantized in units of 2e$^2$/h, which is twice the number reported in earlier studies \cite{InAs1Dwire1,InAs1Dwire2, InAs1Dwire3, InAs1Dwire4,InAscleanwire}. This finding underscores the high material quality and indicates ballistic transport in the gate-defined 1D QPCs.
The observed conductance trace is consistent across different cooldown cycles, but it shows a gradual shift toward more negative pinch-off voltages with repeated voltage sweeps. This drift can be attributed to interactions between large applied split gate voltages and charge traps within the dielectric layer \cite{InAscleanwire}. To mitigate this, we employed slightly larger split gates (500 nm in width and length), resulting in a more stable pinch-off voltage and minimal drift - suggesting dielectric charging as a primary cause of the shift.

\subsection{Calculation of the in-plane effective g-factor}
The effective g-factor, $g^*$, was estimated by analyzing the Zeeman splitting under an applied in-plane magnetic field, $B$, and the subband spacing by applying a DC source-drain bias voltage, $V_{sd}$, using the expression \cite{YilmazDC}

\begin{equation}
    |g^*| = \frac{1}{\mu_B}\frac{d(\Delta E_z)}{dB} = \frac{e}{\mu_B}\frac{\delta V_{sd}}{\delta V_{sg}}\frac{\delta V_{sg}}{\delta B},
    \label{g}
\end{equation}
where $\mu_B$ is the Bohr magneton and $E_z$ represents the Zeeman energy.

\begin{figure}[b]
    \includegraphics[width=.5\textwidth]{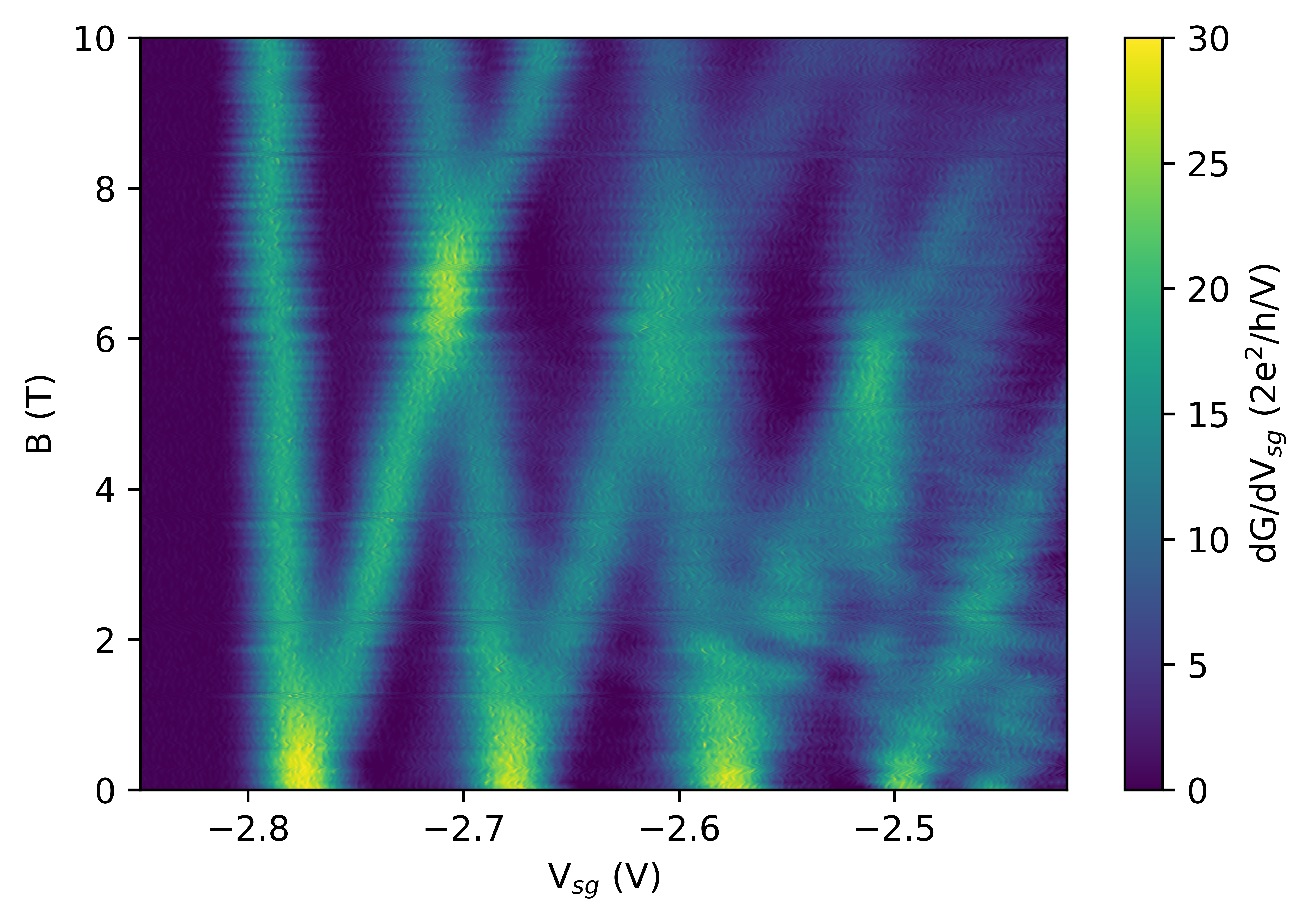} %[width=1\textwidth]

    \caption{Transconductance (d$G$/d$V_{sg}$) plotted as a function of in-plane $B$ on the y-axis and $V_{sg}$ on the x-axis. The bright yellow lines correspond to rises in conductance, while the blue dark regions signify the quantized plateaus. Subband crossings are also visible.}
    \label{b-parallel}
\end{figure}

Clear Zeeman spin-splitting is observed when applying an in-plane magnetic field, as shown in Fig. \ref{b-parallel}. The splitting is visible at approximately 2 T. As the magnetic field increases, odd-numbered plateaus emerge while the even-numbered disappear. At sufficiently high magnetic fields, crossing of the spin-split subbands is observed before the reappearance of the even-numbered plateaus. By extracting the crossover point of the spin-split subbands, $\delta V_{sg}/\delta B$ can be determined. 

\begin{figure}[t]
    \includegraphics[width=.5\textwidth]{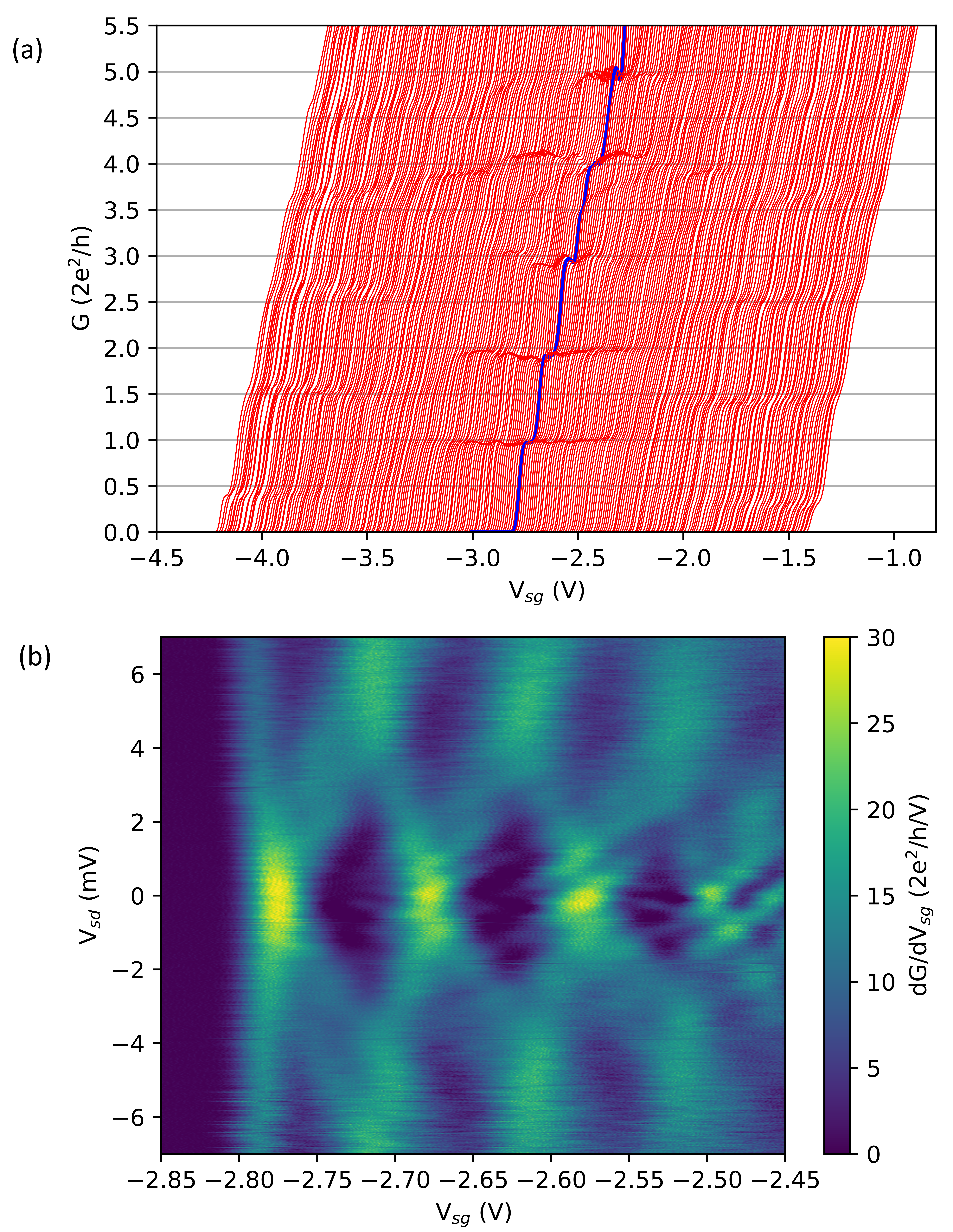} %[width=1\textwidth]

    \caption{(a) $G$ as a function of $V_{sg}$ for different values of $V_{sd}$ varying from - 7 mV (left) to 7 mV (right) in 0.05 mV steps. The blue trace corresponds to $V_{sd}$ = 0 V. For display purposes, the traces have been shifted laterally by 10 mV. The series resistance for this measurement was 1.6 k$\Omega$. (b) Transconductance (d$G$/d$V_{sg}$) plotted as a function of $V_{sd}$ on the y-axis and $V_{sg}$ on the x-axis. The blue dark regions correspond to the quantized plateaus, while the brighter yellow regions indicate the subband transitions where d$G$/d$V_{sg}$ is large.}
    \label{dc}
\end{figure}

Fig. \ref{dc}(a) depicts 1D conductance traces under varying DC bias voltages. When a high DC bias is applied, the momentum degeneracy is lifted \cite{TMChen2} and the spin-degenerate plateaus evolve into 0.5(2e$^2$/h), 1.5(2e$^2$/h), 2.5(2e$^2$/h), etc. For negative DC bias, this behavior is apparent, however, for positive DC bias, a distinct plateau at 0.25(2e$^2$/h) can be seen. This effect has been observed before in InGaAs \cite{YilmazDC} and GaAs \cite{TMChen2} heterostructures, where the asymmetric behavior was attributed to the formation of a density wave or Skyrmion \cite{YilmazDC}. In the case of GaAs \cite{TMChen2}, the emergence of the 0.25(2e$^2$/h) structure was linked to the creation of a unidirectional spin-polarized current. The transconductance as a function of $V_{sd}$ and $V_{sg}$ in Fig. \ref{dc}(b) shows how the even diamonds evolve into odd-quantized values. The intersection where two subbands cross can be used to extract $\delta V_{sd}/\delta V_{sg}$. The effective g-factor of the material was then determined using Eq. \ref{g}. The variables necessary for this calculation are provided in Table \ref{parameters}. The computed effective g-factor is of similar order to that reported in previous studies on InAs QWs \cite{InAscleanwire}.

\begin{table}[h]
\caption{\label{parameters}%
Extracted parameters for the Zeeman splitting and subband spacing crossings used to determine $g^*$.}
\begin{ruledtabular}
\begin{tabular}{ccc}
Subband    &    $N = 1$    &    $N = 2$ \\
$\delta V_{sg}/\delta B (10^{-2}$ V/T)    &    $1.21 \pm 0.06$    &    $1.43 \pm 0.11$\\
$\delta V_{sd}/\delta V_{sg}$ (mV/V)  &    $52.2 \pm 6.6$     &    $43.8 \pm 5.5$  \\
$g^*$    &    $- 10.9 \pm 1.5$    &    $- 10.8 \pm 1.6$ \\
\end{tabular}
\end{ruledtabular}
\end{table}

\subsection{Non-magnetic fractional conductance quantization}
The observation of non-magnetic fractional quantization of conductance typically occurs at low carrier densities and asymmetric confining potentials \cite{Yilmazpaper,Leipaper,Sanjeevpaper}. To investigate fractional conductance quantization in the 1D InAs QPCs, the carrier density was reduced by applying a negative voltage to the top gate. As the top gate voltage became increasingly negative, a sudden transition in conductance was observed in the ground state, shifting from 2e$^2$/h to 2(2e$^2$/h). This transition preceded the emergence of fractional conductance plateaus and may suggest the formation of a zigzag incipient Wigner crystal \cite{KumarMany}.

\begin{figure}[t]
    \includegraphics[width=.5\textwidth]{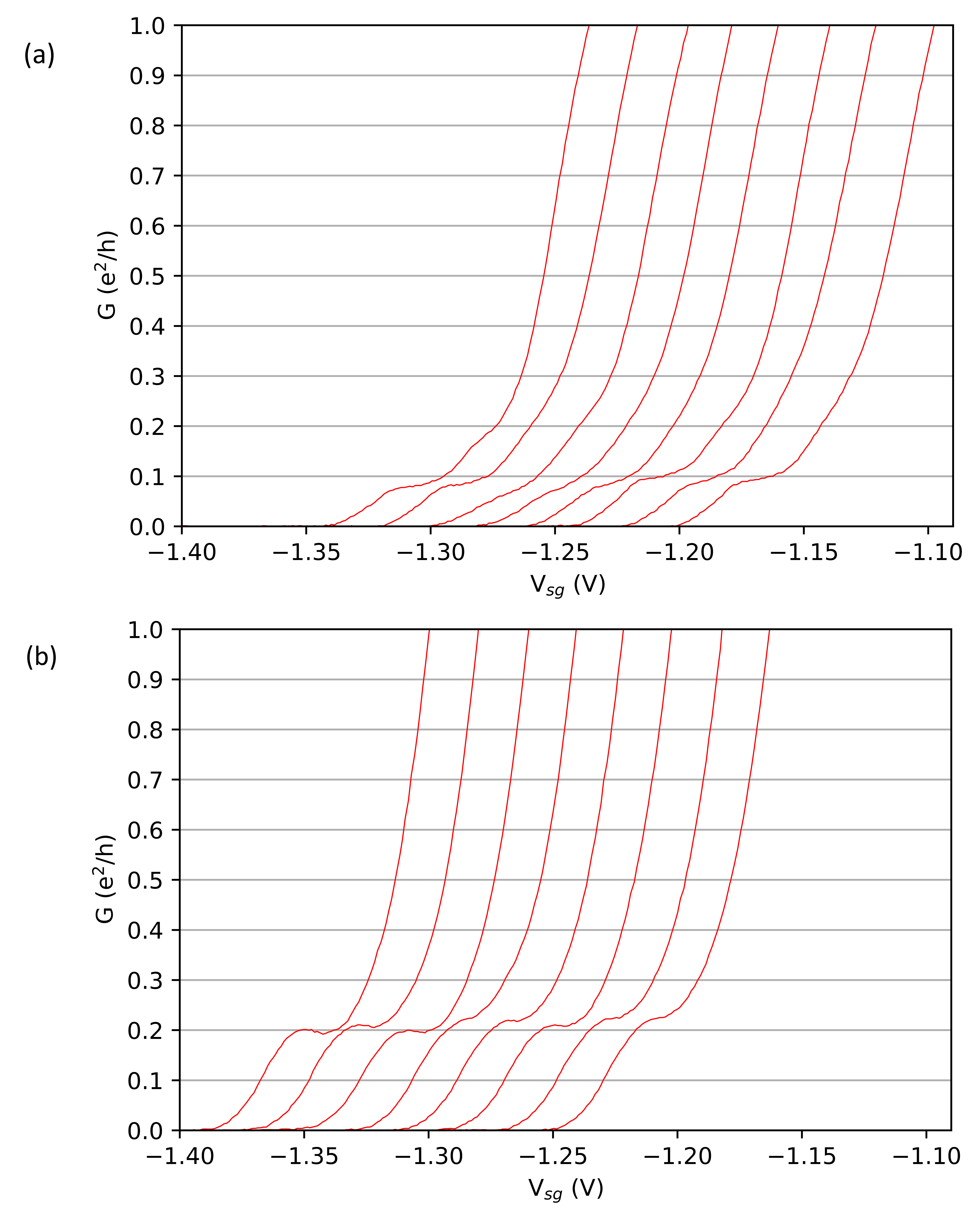} %[width=1\textwidth]

    \caption{Temperature dependence of the (a) 0.1(e$^2$/h) and (b) 0.2(e$^2$/h) plateaus. The 0.1(e$^2$/h) plateau was observed at $V_{tg} = - 3.03$ V, while the 0.2(e$^2$/h) plateau was observed at $V_{tg} = - 2.93$ V. The temperature was varied from 220 mK (left) up to 920 mK (right) in 100 mK steps. For clarity, the traces have been shifted horizontally by 20 mV.}
    \label{temp}
\end{figure}

The 0.1 and 0.2 (e$^2$/h) fractions, previously observed in InGaAs \cite{Leipaper}, are seen in InAs at low carrier densities. Both fractional states persist even as the temperature increases, as shown in Fig. \ref{temp}. Combined with multiple cooldown measurements, this suggests that these non-magnetic fractional states are not caused by disorder. The slight lifting of the plateaus from their quantized value with temperature is consistent with predictions by G. Shavit and Y. Oreg \cite{OregShavit}. Their theoretical model allows for non-magnetic fractional conductance quantization through momentum-conserving backscattering of electrons in two distinct spin subbands or transverse channels, such as those formed in a zigzag incipient Wigner crystal. The resulting conductance is given by

\begin{equation}
    G = \frac{(n_1-n_2)^2}{n_1^2+n_2^2}\left(\frac{e^2}{h}\right),
    \label{Gfrac}
\end{equation}
where $n_i$ represents the number of backscattered electrons from subband $i$. This model predicts the 0.2 (e$^2$/h) plateau when $n_1=1,2$ and $n_2=2,4$, but it does not explain the formation of the 0.1 (e$^2$/h) plateau. A scenario where channel 1 is decoupled from the leads but confined within the wire and strongly interacts with the other channel could yield a conductance of $G=n_1^2/(n_1^2+n_2^2)$ \cite{OregShavit}, potentially explaining the 0.1 (e$^2$/h) state for $n_1=1$ and $n_2=3$. However, no evidence of channel 1 being confined within the wire is observed in this study. Further investigations are necessary to elucidate the origin of the 0.1 (e$^2$/h) plateau.

Applying a magnetic field to these fractional states reveals deviations from previous observations in InGaAs systems \cite{Leipaper}. In InGaAs, the fractions persisted under the application of an in-plane magnetic field parallel to the current direction but disappeared when the field was applied in-plane and perpendicular to the current. In contrast, in InAs, the fractions vanish under an in-plane magnetic field regardless of its orientation relative to the current. 

In InGaAs, this behavior was attributed to Rashba spin splitting. When the magnetic field was parallel to the current, a spin gap formed, enabling backscattering between the spin-split subbands. Conversely, when the magnetic field was perpendicular, one of the spin-split subbands shifted up in energy, preventing backscattering and thereby causing the fractions to disappear. However, the distinct behavior observed in our InAs devices suggests that these states are not spin polarized and do not arise from momentum-conserving backscattering between electrons in different spin subbands. 

This difference can be explained by examining the wafer structures. The InGaAs wafers are Si-doped, which enhances Rashba asymmetry across the QW, while the InAs wafers are undoped. This doping alters the shape of the confining potential and, consequently, the Rashba splitting, leading to the observed variations in behavior.

If a zigzag incipient Wigner crystal has formed, momentum-conserving backscattering of electrons can occur between the two transverse channels of the zigzag. Consequently, the two zigzag channels could become entangled \cite{OregShavit}. 

\begin{figure}[h]
    \includegraphics[width=.5\textwidth]{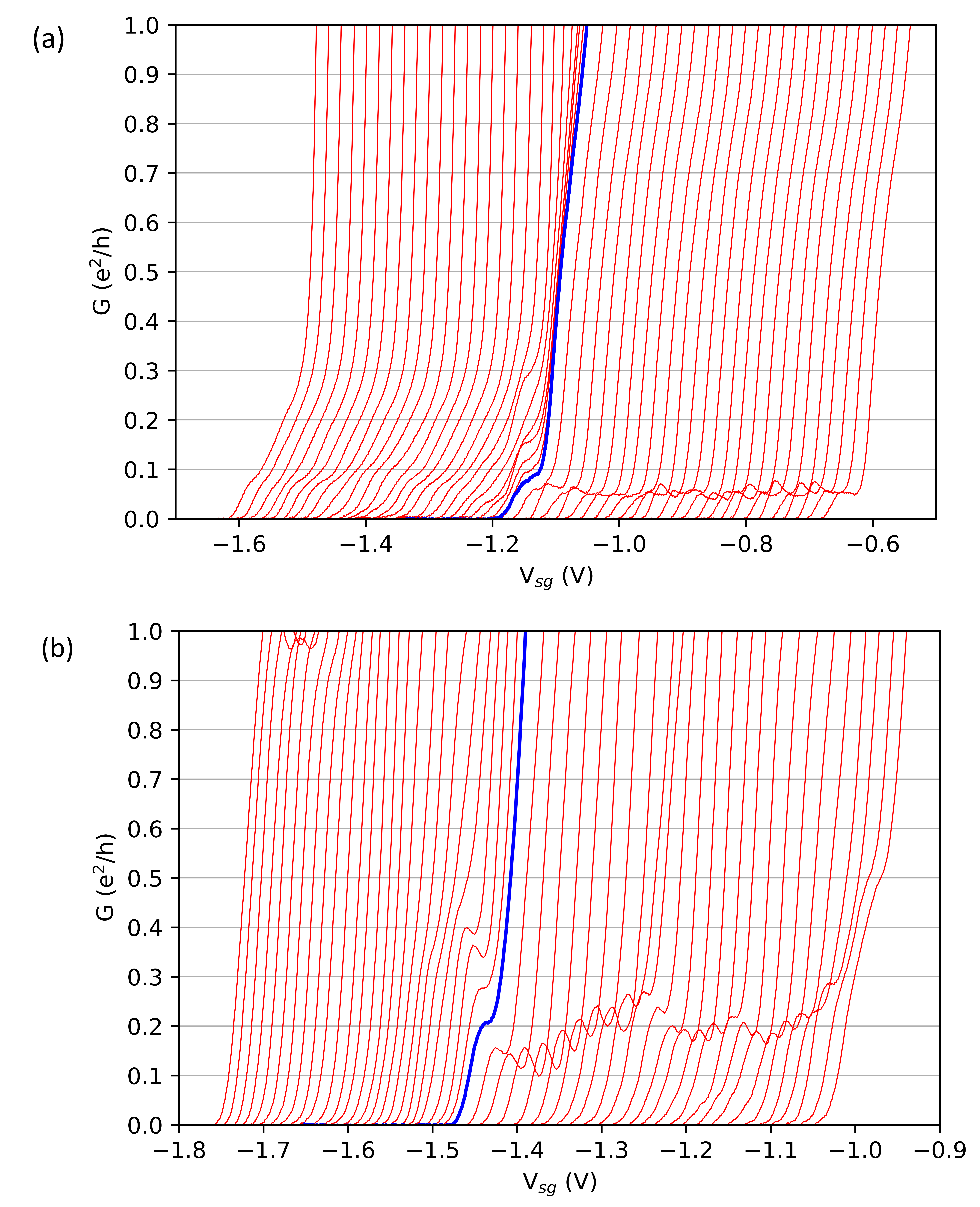} %[width=1\textwidth]
    
    \caption{$G$ as a function of $V_{sg}$ with varying degrees of lateral asymmetric confinement, reaching a maximum shift of $\Delta V_{sg} = \pm 0.5$ V on each side of the graph, for the (a) 0.1(e$^2$/h) and (b) 0.2(e$^2$/h) plateaus. The blue trace corresponds to symmetrically biased split gates. For the 0.1(e$^2$/h) plateau, its value decreases in the direction of the preserved bias, while for the 0.2(e$^2$/h) plateau, it oscillates around its nominal value.}
    \label{lateralasym}
\end{figure}

Fig. \ref{lateralasym} shows the conductance under varying lateral asymmetric potentials, $\Delta V_{sg}$, with fraction formation preserved in one bias direction but suppressed in the opposite. This suggests that one direction of lateral asymmetry may enhance electron-electron interactions, promoting zigzag formation, while the opposite direction weakens these interactions, returning the electrons to their 1D arrangement. Further experiments, such as transverse electron focusing, are needed to conclusively identify zigzag formation.

An interesting aspect of these fractional electron states is their potential to demonstrate fractional charge. Numerous studies \cite{shotnoise-1/3,shotnoise-2/3,shotnoise-nature,shotnoise-nature-1/5,shotnoise-nature-5/2} have investigated fractional charge in fractional quantum Hall effect (FQHE) systems through shot noise experiments. The detection of fractional charge in these InAs fractional conductance states could provide new opportunities to investigate whether they exhibit non-Abelian properties, potentially playing a crucial role in the development of topological quantum computing.

\section{Conclusion} \label{conclusions}
In summary, we have demonstrated high mobility ballistic transport in 1D InAs QPCs, with conductance quantization observed up to 17(2e$^2$/h). The in-plane effective g-factor was calculated to be $- 10.9 \pm 1.5$ for $N=1$ and $- 10.8 \pm 1.6$ for $N=2$, determined from the intersection of spin-split and momentum-split subbands.

Additionally, we have provided evidence for non-magnetic fractional quantization of conductance. Our experiments demonstrate that these states persist at elevated temperatures, are not spin-polarized, and their occurrence is preserved under one direction of lateral asymmetric bias while being suppressed in the opposite direction. This provides experimental support for the theoretical model presented in Ref. \cite{OregShavit}, which suggests that non-magnetic fractional conductance quantization can arise from strong electron-electron interactions and momentum-conserving backscattering between electrons in distinct subbands. 

The data further suggests the potential formation of a zigzag incipient Wigner crystal, where backscattering between the two transverse channels could result in fraction formation and entanglement. Further investigation is required to elucidate the origin of these fractions and their associated subband configurations.

The unique properties of InAs combined with these fractional states suggest promising future applications in spintronic devices and quantum computing systems.

\begin{acknowledgments}
The authors gratefully acknowledge financial support from the UK Science and Technology Facilities Council (Award No. ST/Y005074/1) and the EPSRC (Grant No. EP/R029075/1), which enabled the transport measurements, and the US Department of Energy (Award No. DESC0019274), which supported the growth and characterization efforts at UCSB. We also thank for the use of the shared facilities of the NSF Materials Research Science and Engineering Center (MRSEC) at the University of California Santa Barbara (Grant No. DMR 2308708), and the UCSB Nanofabrication Facility, an open access laboratory, and the SP Cleanroom at the University of Cambridge for access to their fabrication facilities.
\end{acknowledgments}

\bibliography{apssamp}% Produces the bibliography via BibTeX.

\end{document}